\documentclass{svproc}
\usepackage{url}
\usepackage{amsmath}
\usepackage{amssymb}
\usepackage{graphicx}
\usepackage{subcaption}
\usepackage{xcolor}
\usepackage{array}
\usepackage{pifont}
\usepackage{adjustbox}
\newcommand{\PreserveBackslash}[1]{\let\temp=\\#1\let\\=\temp}
\usepackage{multirow}
\newcolumntype{C}[1]{>{\PreserveBackslash\centering}p{#1}}
\newcolumntype{R}[1]{>{\PreserveBackslash\raggedleft}p{#1}}
\newcolumntype{L}[1]{>{\PreserveBackslash\raggedright}p{#1}}
\begin{document}
\mainmatter
%

\title{DCD: A Semantic Segmentation Model for Fetal Ultrasound Four-Chamber View}
\titlerunning{DCD: A Semantic Segmentation Model}  
%
\author{Donglian Li\inst{1} \and Hui Guo\inst{2,3,*}\and 
Minglang Chen\inst{4,3} \and Huizhen Chen\inst{5}\and Jialing Chen\inst{6} \and Bocheng Liang\inst{7}\and Pengchen Liang\inst{8} \and Ying Tan\inst{7}}
\authorrunning{Donglian Li et al.} 
%
\tocauthor{Ivar Ekeland, Roger Temam, Jeffrey Dean, David Grove,
Craig Chambers, Kim B. Bruce, and Elisa Bertino.}
\institute{School of Information and Communication, Guilin University of Electronic Technology, Guilin 541004, China
\and
Faculty of Humanities and Arts, Macau University of Science and Technology, Macau 999078, China\\
\email{3220002921@student.must.edu.mo}
\and
Guangxi Key Laboratory of Machine Vision and Intelligent Control, Wuzhou University, Wuzhou 543002, China
\and
Faculty of Innovation Engineering, Macau University of Science and Technology, Macau 999078, China
\and
Department of Public Education, Wuzhou Medical College, Wuzhou 543199, China
\and
College of Big Data and Software Engineering, Wuzhou University, Wuzhou 543002, China
\and
Shenzhen Maternity and Child Healthcare Hospital, Southern Medical University, Shenzhen 518100, China
\and
School of Microelectronics, Shanghai University, Shanghai 201800, China
}
\maketitle              

\begin{abstract}
Accurate segmentation of anatomical structures in the apical four-chamber (A4C) view of fetal echocardiography is essential for early diagnosis and prenatal evaluation of congenital heart disease (CHD). However, precise segmentation remains challenging due to ultrasound artifacts, speckle noise, anatomical variability, and boundary ambiguity across different gestational stages. To reduce the workload of sonographers and enhance segmentation accuracy, we propose DCD, an advanced deep learning-based model for automatic segmentation of key anatomical structures in the fetal A4C view. Our model incorporates a Dense Atrous Spatial Pyramid Pooling (Dense ASPP) module, enabling superior multi-scale feature extraction, and a Convolutional Block Attention Module (CBAM) to enhance adaptive feature representation. By effectively capturing both local and global contextual information, DCD achieves precise and robust segmentation, contributing to improved prenatal cardiac assessment.

\keywords{Apical Four-chamber view, Anatomical Structure Segmentation, Congenital Heart Disease, Dense ASPP, Convolutional Block Attention Module}
\end{abstract}
\section{Introduction}
Congenital Heart Disease (CHD) is one of the most common congenital anomalies and has become the leading cause of birth defects in newborns, making it a significant contributor to infant mortality~\cite{In1}. 
Severe CHD can severely disrupt fetal and neonatal circulation, and if not promptly diagnosed and treated, it may lead to serious cardiac dysfunction and even pose a life-threatening risk~\cite{In2}.
Therefore, the early and accurate diagnosis of CHD is essential for facilitating timely and effective medical interventions, improving treatment outcomes, and ultimately enhancing the survival and quality of life of affected infants~\cite{In3}.
Echocardiography, with its non-invasiveness, cost-effectiveness, and real-time imaging capabilities, has become the preferred method for prenatal CHD screening and plays a crucial role in the early detection of cardiac structural abnormalities~\cite{In4}.
During prenatal screening, sonographers typically capture multiple critical fetal echocardiographic images to assess cardiac development and detect potential abnormalities. Among these, the Apical Four-Chamber (A4C) view is considered one of the most essential imaging planes, as it serves as both the primary and earliest approach for detecting congenital heart disease (CHD) in fetuses~\cite{In5}. Direct observation of the A4C view allows for the preliminary screening of certain basic fetal cardiac conditions. However, the assessment of more complex CHD cases requires further evaluation through organ segmentation and precise parameter measurements to ensure a more comprehensive and accurate diagnosis.
Traditionally, segmentation of anatomical structures in the A4C view was performed manually by sonographers, requiring substantial time and effort. Moreover, sonographers require extensive professional knowledge and long-term practical experience to perform segmentation tasks as accurately as possible. Noise, artifacts, and the dynamic motion of the fetal heart in ultrasound imaging often blur anatomical structures and obscure boundaries. This makes manual delineation susceptible to subjective bias, potentially compromising diagnostic accuracy. 
Developing an automated segmentation method is essential for improving efficiency in fetal echocardiographic analysis, reducing the workload of sonographers, and advancing intelligent ultrasound diagnostics in clinical practice.
We propose a semantic segmentation model for the A4C view, named DCD, where the first D represents DeepLabv3+, C denotes the Convolutional Block Attention Module (CBAM), and the final D stands for Dense Atrous Spatial Pyramid Pooling (Dense ASPP).
This model enables the automatic segmentation of anatomical structures within the A4C view and demonstrates robustness across various challenging imaging conditions. 
The main contributions of this paper are summarized as follows:
\begin{itemize}
\item We propose a semantic segmentation model specifically designed for the A4C view, capable of automatically identifying and segmenting 13 key anatomical structures within the A4C view. Compared to other baseline models, it achieves higher segmentation accuracy.
\item The model integrates the Dense ASPP module, which enhances feature extraction capabilities by minimizing pixel-level feature loss and preserving feature integrity. This improvement significantly strengthens segmentation performance, particularly in delineating boundary regions.
\item The model employs the CBAM to simultaneously refine channel and spatial attention allocation, enhancing the representation of critical features. This optimization effectively reduces segmentation omissions and target misclassifications, thereby significantly improving segmentation accuracy.
\end{itemize}
\section{Related Work}
\subsection{Fetal Ultrasound Image Segmentation}
In recent years, deep learning methods have achieved significant progress in the field of medical ultrasound image segmentation~\cite{R11,R16,R13,R17,R15,R18}. 
Pu et al.~\cite{R12} introduced a deep learning model integrating MobileNet, UNet, and Feature Pyramid Networks (FPN). By leveraging a multi-level edge computing system, their method enables efficient training and enhances the segmentation accuracy of fetal echocardiographic cardiac structures.
Chen et al.~\cite{R13} combined deep convolutional networks, multi-channel Transformers, and cross-layer graph convolution modules to enhance the integration of local, global, and structural information, achieving high-precision segmentation of thyroid nodules. 
Zhao et al.~\cite{R14} incorporated a deformable self-attention mechanism, a boundary-aware decoder, and an auxiliary segmentation head to improve multi-scale information processing and local feature extraction, enabling precise segmentation and biometric measurement of multiple fetal anatomical structures.
Although these studies have achieved significant advancements in medical ultrasound image segmentation, challenges such as complex background interference and weak boundary segmentation remain. Therefore, further exploration of more efficient segmentation methods is necessary to enhance the automation of ultrasound image analysis and facilitate its clinical diagnostic applications.
\subsection{Attention Mechanisms in Medical Image Segmentation}
The attention mechanism is a technique in deep learning that simulates human attention, enabling models to focus on the most relevant parts of the input when processing information. In recent years, attention mechanisms have been widely applied in medical image analysis, demonstrating outstanding performance, particularly in visual tasks such as ultrasound image segmentation~\cite{R21,R22,R23,R24,R25}.  
Rajamani et al.~\cite{R22} introduced a regularization approach by injecting noise into attention blocks, enhancing the robustness of attention blocks and integrating them into the bottleneck layer of U-Net, thereby improving segmentation performance under limited training data conditions.  
Tong et al.~\cite{R24} proposed a hybrid attention mechanism model, where a channel attention enhancement module and a feature optimization module work in synergy to improve the accuracy of multi-organ medical image segmentation. Cao et al.~\cite{R25} refined the U-Net architecture by incorporating a context extraction module, spatial and channel attention-based skip connections, and a novel loss function, resulting in improved segmentation accuracy.  
In summary, attention mechanisms have achieved remarkable progress in medical image segmentation. In this study, we incorporate the CBAM attention mechanism to enhance channel and spatial attention interactions, optimize feature representation, and further improve segmentation accuracy.

\section{Methodology}
\subsection{DCD Model}

\begin{figure}[htbp]
\centerline{\includegraphics[width=\textwidth]{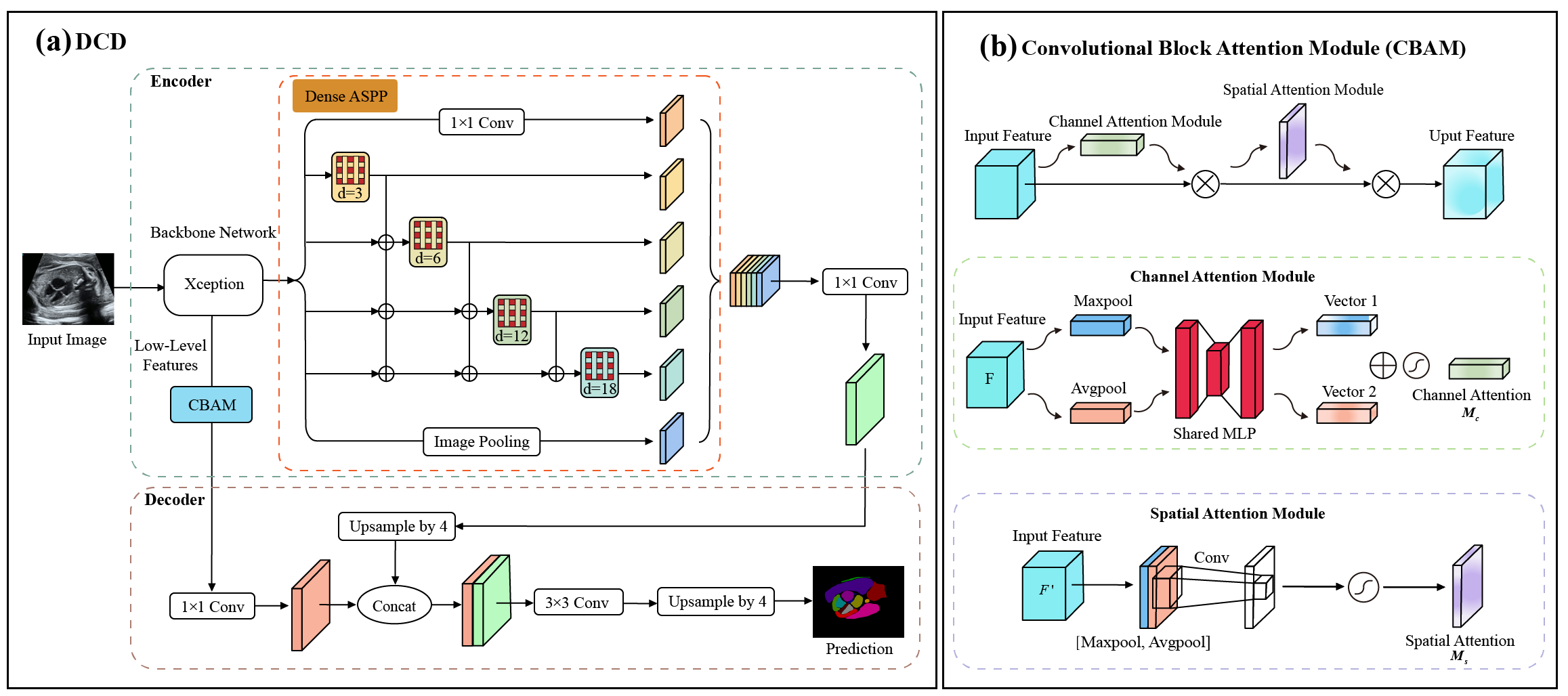}}
\caption{An overview of the proposed DCD model.}
\label{fig1}
\end{figure}

We propose a novel improved semantic segmentation model based on the DeepLabv3+~\cite{DeepLabv3+} architecture, with its framework illustrated in Fig.~\ref{fig1}. The model adopts an encoder-decoder structure, where the encoder utilizes Xception as the backbone network. During feature extraction, shallow features are enhanced by incorporating the CBAM attention mechanism to improve local feature representation, while deep features leverage Dense ASPP to expand the receptive field.  
In the decoder, high-resolution features are progressively restored to enhance the model's ability to recognize target regions and improve segmentation accuracy. This model is capable of segmenting 13 key anatomical structures in the A4C view, with the detailed anatomical information presented in Table~\ref{table1}.
\begin{table}[!ht]
    \centering
    \caption{Professional terms and corresponding abbreviations of anatomical structures.}
    \begin{tabular}{cc|cc}
    \hline
        Abbreviation  & Description  & Abbreviation  & Description  \\ \hline
        SP  & Spine  & VS  & Ventricular Septum  \\ 
        RiB  & Ribs  & LVW  & Left Ventricular Wall  \\ 
        LA  & Left Atrium  & RVW  & Right Ventricular Wall  \\ 
        IS  & Interatrial Septum  & DAO  & Descending Aorta  \\ 
        RA  & Right Atrium  & RL  & Right Lung  \\ 
        RV  & Right Ventricle  & LL  & Left Lung  \\ 
        LV  & Left Ventricle  &   &   \\ \hline
    \end{tabular}
    \label{table1}
\end{table}
\subsection{Dense ASPP}
The ASPP module in the DeepLabv3+ model has insufficient feature resolution, resulting in the degradation of feature representation and potential information loss. Additionally, the discrete nature of parallel dilated convolutions in image space can cause discontinuities in the segmentation of small-scale structures, which in turn affects the precise extraction of detailed anatomical features. To address these challenges, we replace the original ASPP module with Dense ASPP, which enhances the model's ability to extract anatomical structure features in the apical four-chamber view. Dense ASPP then fuses features from different scales and transmits them to the next layer via dense connections. This enables Dense ASPP to efficiently leverage multi-scale features while effectively processing both fine details and contextual information. The detailed architecture of Dense ASPP is shown in Fig.~\ref{fig1}(a), which includes convolutional layers with dilation rates of 3, 6, 12, and 18, allowing for parallel connections of features at four distinct scales. The computation process of Dense ASPP is expressed in Eq.~(\ref{eq11}).
\begin{equation}
 F_l = f_0\left(\text{Conv}_{K,d_l}\left([F_0, F_1, \dots, F_{l-1}]\right)\right),
  \label{eq11}
\end{equation}
where \( F_l \) represents the output feature of the \( l \)-th layer, \( \text{Conv}_{K, d_l}(\cdot)
 \) denotes the dilated convolution operation with a kernel size of \( K \) and dilation rate \( d_l \), \( [F_0, F_1, \dots, F_{l-1}] \) represents the input of the current layer, which is formed by concatenating the output features from all preceding layers, and \( f_0 (\cdot)\) represents the nonlinear transformation.

The traditional ASPP has dilation rates of 6, 12, and 18, resulting in receptive fields of 13, 25, and 37, respectively. In contrast, the Dense ASPP structure enhances feature information reuse across different layers, thereby increasing the receptive field. For instance, when the branches with dilation rates of 6 and 12 are connected, the receptive field increases to 37, which is the same as the maximum receptive field of the traditional ASPP. This demonstrates that Dense ASPP effectively expands the receptive field through inter-layer information reuse, thereby improving the ability to extract multi-scale features.

\subsection{Convolutional Block Attention Module}
The CBAM attention mechanism, composed of a channel attention module and a spatial attention module, enhances feature representation in deep learning models. The channel attention module focuses on assigning importance weights to feature map channels, while the spatial attention module enhances the model’s ability to highlight distinctive pixel regions within an image, as illustrated in Fig.~\ref{fig1}(b). The computational processes of channel attention and spatial attention in the CBAM attention mechanism can be mathematically represented by Eq.~(\ref{eq1}) and Eq.~(\ref{eq2}).
\begin{eqnarray}
F^{\prime}=M_c(F)\otimes F , \label{eq1} \\
F^{\prime\prime}=M_s(F^{\prime})\otimes F^{\prime} , \label{eq2}
\end{eqnarray}
where \( F \) represents the feature map, and \( F' \) and \( F'' \) denote the feature maps output by the channel attention module and spatial attention module, respectively, where \( M_c \) and \( M_s \) represent the channel and spatial attention mechanisms, respectively.

The channel attention module primarily focuses on learning the correlation between feature channels and performs weighted fusion of different channel features. First, global average pooling (GAP) and global maximum pooling (GMP) are applied to the input feature map \( F \in \mathbb{R}^{C \times H \times W} \) to compute the statistical information along the channel dimension, as shown in Eq.~(\ref{eq3}) and Equation Eq.~(\ref{eq4}).
\begin{eqnarray}
F_{\text{avg}}^c = \frac{1}{H \times W} \sum_{i=1}^{H} \sum_{j=1}^{W} F(c,i,j), \label{eq3} \\
F_{\text{max}}^c = \max_{i \in H, j \in W} F(c,i,j)
 , \label{eq4}
\end{eqnarray}
where \( F_{\text{avg}}^c \) and \( F_{\text{max}}^c \) represent the results of global average pooling and global maximum pooling, respectively. The two pooling results are then subjected to a nonlinear transformation through a shared MLP, which can be expressed as Eq.~(\ref{eq5}).

\begin{equation}
 M_c = \delta\left( W_2\left( \delta\left( W_1\left( F_{\text{avg}}  \right) \right) \right) + W_2\left( \delta\left( W_1\left( F_{\text{max}} \right) \right) \right)\right)
,
  \label{eq5}
\end{equation}
where \( W_1 \in \mathbb{R}^{C/r \times C} \) and \( W_2  \in \mathbb{R}^{C \times C/r} \)  are the parameters of the two fully connected layers, \( r \) is the channel scaling coefficient, \( \delta(\cdot) \) is the ReLU activation function, and \( \sigma(\cdot) \) is the Sigmoid activation function, which ensures that the output values lie between 0 and 1. The final weighted channel attention feature is obtained as shown in Eq.~(\ref{eq6}).
\begin{equation}
F' = M_c \cdot F
,
  \label{eq6}
\end{equation}
where \( M_c \) is the computed channel attention weight, and \( F \) is the input feature map.

The spatial attention module primarily focuses on learning the importance of different regions in the image, performing weighted fusion of information from different spatial locations. First, global maximum pooling and global average pooling are applied to the input feature map \( F' \in \mathbb{R}^{C \times H \times W} \) along the channel dimension, as shown in Eq.~(\ref{eq7}) and Eq.~(\ref{eq8}).
\begin{eqnarray}
F_{\text{avg}}^s = \frac{1}{C} \sum_{c=1}^{C} F'(c,i,j)
, \label{eq7} \\
F_{\text{max}}^s = \max_{c \in C} F'(c,i,j)
, \label{eq8}
\end{eqnarray}
where \( F_{\text{avg}}^s, F_{\text{max}}^s \in \mathbb{R}^{1 \times H \times W}
 \) represent the feature maps. These two feature maps are concatenated and then a \( 7 \times 7 \) convolution operation is applied to compute the spatial attention weight, as shown in Eq.~(\ref{eq9}):
\begin{equation}
M_s = \delta\left(\text{Conv}_{7 \times 7}\left([F_{\text{avg}}^s, F_{\text{max}}^s]\right)\right)
,
  \label{eq9}
\end{equation}
where \( [F_{\text{avg}}^s, F_{\text{max}}^s] \) represents the concatenation along the channel dimension. The weighted spatial attention feature is then computed, as shown in Eq.~(\ref{eq10}):
\begin{equation}
F'' = M_s \cdot F'
,
  \label{eq10}
\end{equation}
where \( M_s \) is the computed spatial attention weight, and \( F' \) is the feature map after channel attention.
\subsection{Loss Function}
This study employs a combination of Cross Entropy Loss (CE) and Dice Similarity Coefficient (DSC) loss functions as the objective functions.
The Cross Entropy Loss measures the discrepancy between the predicted probability distribution and the true class distribution. By minimizing this loss, the discrepancy between the predicted and target values is reduced. As the loss decreases, the model's predictions become increasingly closer to the true classes, resulting in improved model performance. The expression for the CE loss is given in Equation in Eq.~(\ref{eq12}).
\begin{equation}
L_{\text{CE}} = -\frac{1}{n} \sum_{i=1}^{n} \sum_{c=1}^{C} y_{i,c} \ln p_{i,c}
,
  \label{eq12}
\end{equation}
where \( n \) represents the batch size, \( C \) denotes the number of classes, \( y_{i,c} \) indicates the ground truth label of sample \( i \) for class \( c \), and \( p_{i,c} \) represents the predicted probability of the model for class \( c \). By optimizing the cross-entropy loss, the model's classification accuracy for multi-class segmentation tasks can be improved.

The Dice loss function is a set similarity measure used to calculate the similarity between two samples. The expression for the Dice loss function is given in Eq.~(\ref{eq13}).
\begin{equation}
L_{\text{Dice}} = 1 - \frac{2|X \cap Y|}{|X| + |Y|},
  \label{eq13}
\end{equation}
where \( X \) represents the predicted segmentation result, \( Y \) represents the ground truth segmentation label, and \( |X \cap Y| \) denotes the intersection of the predicted segmentation result and the ground truth label, i.e., their overlapping region. \( |X| \) is the total number of pixels in the predicted segmentation result, and \( |Y| \) is the total number of pixels in the ground truth label. Therefore, the overall loss function is expressed as Eq.~(\ref{eq14}):
\begin{equation}
L = L_{\text{CE}} + L_{\text{Dice}}
.
  \label{eq14}
\end{equation}

\section{Experiments and Results}
\subsection{Dataset and Implementation Details}
This study collected prenatal ultrasound data from Shenzhen Maternal and Child Healthcare Hospital. After complete anonymization under the guidance of the relevant departments in the hospital, the data was used for research purposes, with approval number LLYJ2022-014-005. The data collection equipment includes various brands such as Siemens, Philips, and Samsung, and all ultrasound images were annotated by experienced sonographer. Our dataset contains a total of 1,152 fetal apical four-chamber heart images, which are split into training, validation, and test sets in a 7:2:1 ratio.

All experiments in this paper were conducted on the Ubuntu 20.04 operating system, with the experimental environment being Python 3.8, PyTorch 1.13, and CUDA 11.7. The hardware configuration includes an Intel(R) Xeon(R) Platinum 8375C CPU @ 2.90GHz and an NVIDIA GeForce RTX 3050 6GB Laptop GPU. The optimizer used is the Adam optimizer with momentum, a momentum value of 0.9. The input image size is set to 512×512, with a batch size of 8 and 400 training epochs. The minimum and maximum learning rates are set to \(5 \times 10^{-6}\) and \(5 \times 10^{-4}\), respectively, with the learning rate decay method set to cosine annealing.
\subsection{Evaluation Metric}
This study employs Intersection over Union (IoU) as the evaluation metric to comprehensively assess the performance of the segmentation methods. IoU measures the overlap between the ground truth and the predicted output for each pixel class, and is defined as shown in Eq.~(\ref{eq15}).
\begin{equation}
\text{IoU} = \frac{|G_c \cap P_c|}{|G_c \cup P_c|}
,
  \label{eq15}
\end{equation}
where \( G_c \) represents the set of pixels corresponding to the true class \( c \), \( P_c \) represents the set of pixels predicted by the model as class \( c \), \( |G_c \cap P_c| \) denotes the intersection of the true and predicted pixels for class \( c \), and \( |G_c \cup P_c| \) represents the union of the true and predicted pixels for class \( c \).
\subsection{Ablation Experiment}
\begin{table}[!ht]
    \centering
    \caption{Effect of different modules on segmentation performance.}
    \begin{tabular}{ccc|ccccccc|c}
    \hline
        Dense ASPP & ASPP & CBAM & RiB  & SP  & RA  & RV  & LA  & LV  & VS  & mIoU \\ \hline
        \multirow{3}{*}{\ding{55}} & \multirow{3}{*}{\checkmark} & \multirow{3}{*}{\ding{55}} & 55.41   & 78.67   & 88.22   & 80.04   & \underline{83.61}   & 81.94   & 71.71   & \multirow{3}{*}{74.95} \\ 
        & ~ & ~ & LVW  & RVW  & IS  & DAO  & RL  & LL  &   & ~ \\ 
        & ~ & ~ & 64.71   & 66.64   & 53.33   & 76.00   & 87.17   & 86.92   &   & ~ \\ \hline
        \multirow{4}{*}{\ding{55}} & \multirow{4}{*}{\checkmark} & \multirow{4}{*}{\checkmark} & RiB  & SP  & RA  & RV  & LA  & LV  & VS  & \multirow{4}{*}{75.31} \\ 
        ~ & ~ & ~ & \underline{55.85}   & 78.40   & \underline{88.62}   & \underline{81.28}   & 83.24   & 81.93   & 72.38   & ~ \\ 
        ~ & ~ & ~ & LVW  & RVW  & IS  & DAO  & RL  & LL  &   & ~ \\ 
        ~ & ~ & ~ & 66.71   & \underline{67.72}   & 53.39   & 75.73   & 86.99   & 86.79   &   & ~ \\ \hline
        \multirow{4}{*}{\checkmark} & \multirow{4}{*}{\ding{55}} & \multirow{4}{*}{\ding{55}} & RiB  & SP  & RA  & RV  & LA  & LV  & VS  & \multirow{4}{*}{\underline{75.73}} \\ 
        ~ & ~ & ~ & 55.43   & \underline{78.84}  & \textbf{88.98}   & \textbf{82.16}   & \textbf{83.80}   & \underline{82.68}   & \textbf{73.58}   & ~ \\ 
        ~ & ~ & ~ & LVW  & RVW  & IS  & DAO  & RL  & LL  &   & ~ \\ 
        ~ & ~ & ~ & \underline{67.00}   & 67.63   & \underline{53.43}   & \underline{76.66}   & \underline{87.20}   & 87.15   &   & ~ \\ \hline
        \multirow{4}{*}{\checkmark} & \multirow{4}{*}{\ding{55}} & \multirow{4}{*}{\checkmark} & RiB  & SP  & RA  & RV  & LA  & LV  & VS  & \multirow{4}{*}{\textbf{76.01}} \\ 
        ~ & ~ & ~ & \textbf{55.98}   & \textbf{79.44}   & 88.48   & 80.66   & 83.45   & \textbf{82.95}   & \underline{72.47}   & ~ \\ 
        ~ & ~ & ~ & LVW  & RVW  & IS  & DAO  & RL  & LL  &   & ~ \\ 
        ~ & ~ & ~ & \textbf{67.39}   & \textbf{68.42}   & \textbf{55.71}   & \textbf{77.81}   & \textbf{87.57}   & \textbf{87.75}   &   & ~ \\ \hline
    \end{tabular}
    \label{table2}
\end{table}

We conducted ablation experiments on an internal dataset to evaluate the impact of the Dense ASPP structure and the CBAM attention mechanism on the model's performance. The results of the ablation experiments are shown in Table~\ref{table2}. When ASPP was used as the feature extraction module alone, the overall mIoU was 74.95. After incorporating the CBAM attention mechanism into the ASPP structure, the mIoU increased to 75.31, indicating that the attention mechanism contributes to improved segmentation performance. Furthermore, by replacing the traditional ASPP with Dense ASPP, the mIoU was further improved to 75.73, demonstrating that Dense ASPP is more effective at capturing multi-scale features. When Dense ASPP and CBAM were combined, the mIoU reached 76.01, which was the best result across all experimental setups, showing that Dense ASPP and CBAM work cooperatively to enhance the segmentation performance of key fetal heart structures. Moreover, the IoU results for specific structures show that most anatomical structures achieved better segmentation performance with the Dense ASPP + CBAM combination, especially in structures such as SP, LV, and LL, further validating the effectiveness of the proposed method.

\subsection{Comparison Experiment}
To further evaluate the performance of the DCD model, we conducted comparison experiments with several mainstream semantic segmentation models from different years, such as UNet~\cite{UNet}, DeepLabv3+~\cite{DeepLabv3+}, and Mask2Former~\cite{Mask2Former}. As shown in Table~\ref{table3}, DCD demonstrated outstanding performance across multiple key anatomical structure segmentation tasks, achieving the best overall segmentation performance with an mIoU of 76.01. Notably, the model performed exceptionally well in the segmentation of various anatomical structures, such as RiB, SP, and RL. For instance, DCD achieved an IoU of 55.98 in the segmentation of the RiB structure, significantly outperforming other models, indicating its ability to accurately capture complex and fine anatomical structures of the fetal heart. Additionally, the model exhibited significant advantages in the segmentation of SP and LA structures, with IoU of 79.44 and 87.57, respectively, surpassing the baseline models by a considerable margin.

These results demonstrate that DCD is highly effective in capturing and utilizing critical anatomical information from fetal hearts, particularly excelling in the segmentation of challenging anatomical structures. Through these comparative experiments, we have validated the strong performance of DCD in multi-class and complex structure segmentation tasks, further proving the model’s effectiveness and superiority in fetal ultrasound heart image segmentation.

\begin{table}[!ht]
    \centering
    \caption{Performance comparison of different semantic segmentation models.}
    \adjustbox{valign=c, max width=\textwidth}{
    \begin{tabular}{c|>{\centering\arraybackslash}p{1cm} >{\centering\arraybackslash}p{1cm} >{\centering\arraybackslash}p{1cm} >{\centering\arraybackslash}p{1cm} >{\centering\arraybackslash}p{1cm} >{\centering\arraybackslash}p{1cm} >{\centering\arraybackslash}p{1cm}|c}
    \hline
        Method & RiB  & SP  & RA  & RV  & LA  & LV  & VS  & mIoU \\ \hline
        \multirow{3}{*}{FCN~\cite{FCN}} & 53.16   & 78.16   & 87.36   & 80.26   & 82.98   & 82.10   & 70.84   & \multirow{3}{*}{73.84} \\ 
        ~ & LVW  & RVW  & IS  & DAO  & RL  & LL  &   & ~ \\ 
        ~ & \underline{67.05}   & 66.36   & 47.57   & 72.69   & 85.66   & 85.72   &   & ~ \\ \hline
         \multirow{4}{*}{UNet~\cite{UNet}} & RiB  & SP  & RA  & RV  & LA  & LV  & VS  & \multirow{4}{*}{73.29} \\ 
        ~ & 53.32   & 77.03   & 87.56   & 77.22   & 83.42   & 80.43   & 70.18   & ~ \\ 
        ~ & LVW  & RVW  & IS  & DAO  & RL  & LL  &   & ~ \\ 
        ~ & 63.07   & 62.21   & 52.31   & 76.34   & 84.88   & 84.74   &   & ~ \\ \hline
        \multirow{4}{*}{DeepLabv3+~\cite{DeepLabv3+}} & RiB  & SP  & RA  & RV  & LA  & LV  & VS  & \multirow{4}{*}{73.68} \\ 
        ~ & \underline{55.41}   & \underline{78.67}   & 88.22   & 80.04   & 83.61   & 81.94   & 71.71   & ~ \\ 
        ~ & LVW  & RVW  & IS  & DAO  & RL  & LL  &   & ~ \\ 
        ~ & 64.71   & 66.64   & 53.33   & 76.00   & \underline{87.17}   & 86.92   &   & ~ \\ \hline
        \multirow{4}{*}{DANet~\cite{DANet}} & RiB  & SP  & RA  & RV  & LA  & LV  & VS  & \multirow{4}{*}{74.25} \\ 
        ~ & 53.42   & 78.14   & 87.38   & 80.33   & \underline{83.65}   & 81.61   & 71.15   & ~ \\ 
        ~ & LVW  & RVW  & IS  & DAO  & RL  & LL  &   & ~ \\ 
        ~ & 66.53   & 66.36   & 50.87   & 73.48   & 85.80   & 86.52   &   & ~ \\ \hline
        \multirow{4}{*}{FastFCN~\cite{FastFCN}} & RiB  & SP  & RA  & RV  & LA  & LV  & VS  & \multirow{4}{*}{71.76} \\ 
        ~ & 48.41   & 76.92   & 85.70   & 78.14   & 81.21   & 79.85   & 67.06   & ~ \\ 
        ~ & LVW  & RVW  & IS  & DAO  & RL  & LL  &   & ~ \\ 
        ~ & 64.65   & 63.03   & 46.51   & 69.73   & 85.66   & 86.01   &   & ~ \\ \hline
         \multirow{4}{*}{BiSeNet V2~\cite{BiSeNetV2}} & RiB  & SP  & RA  & RV  & LA  & LV  & VS  & \multirow{4}{*}{66.95} \\ 
        ~ & 46.50   & 68.21   & 80.41   & 72.27   & 75.08   & 75.17   & 66.92   & ~ \\ 
        ~ & LVW  & RVW  & IS  & DAO  & RL  & LL  &   & ~ \\ 
        ~ & 58.96   & 58.81   & 36.92   & 65.84   & 82.10   & 83.10   &   & ~ \\ \hline
         \multirow{4}{*}{Mask2Former~\cite{Mask2Former}} & RiB  & SP  & RA  & RV  & LA  & LV  & VS  & \multirow{4}{*}{\underline{75.55} }\\ 
        ~ & 53.38   & 78.01   & \textbf{89.18}   & \textbf{81.46}   & \textbf{83.76}   & \textbf{84.16}   & \textbf{72.94}   & ~ \\ 
        ~ & LVW  & RVW  & IS  & DAO  & RL  & LL  &   & ~ \\ 
        ~ & 66.36  & \underline{68.33}   & 54.20   & 77.14   & 86.35   & \underline{86.84}  &   & ~ \\ \hline
         \multirow{4}{*}{DC-DeepLabv3+ (Ours)} & RiB  & SP  & RA  & RV  & LA  & LV  & VS  & \multirow{4}{*}{\textbf{76.01}} \\ 
        ~ & \textbf{55.98}   & \textbf{79.44}   & \underline{88.48}   & \underline{80.66}   & 83.45   & \underline{82.95}   & \underline{72.47}   & ~ \\ 
        ~ & LVW  & RVW  & IS  & DAO  & RL  & LL  &   & ~ \\ 
        ~ & \textbf{67.39}   & \textbf{68.42}   & \textbf{55.71}   & \textbf{77.81}   & \textbf{87.57}   & \textbf{87.75}   &   & ~ \\ \hline
        
    \end{tabular}
    }
    \label{table3}
\end{table}

\subsection{Visualization Analysis}
\begin{figure}[htbp]
\centerline{\includegraphics[width=\textwidth]{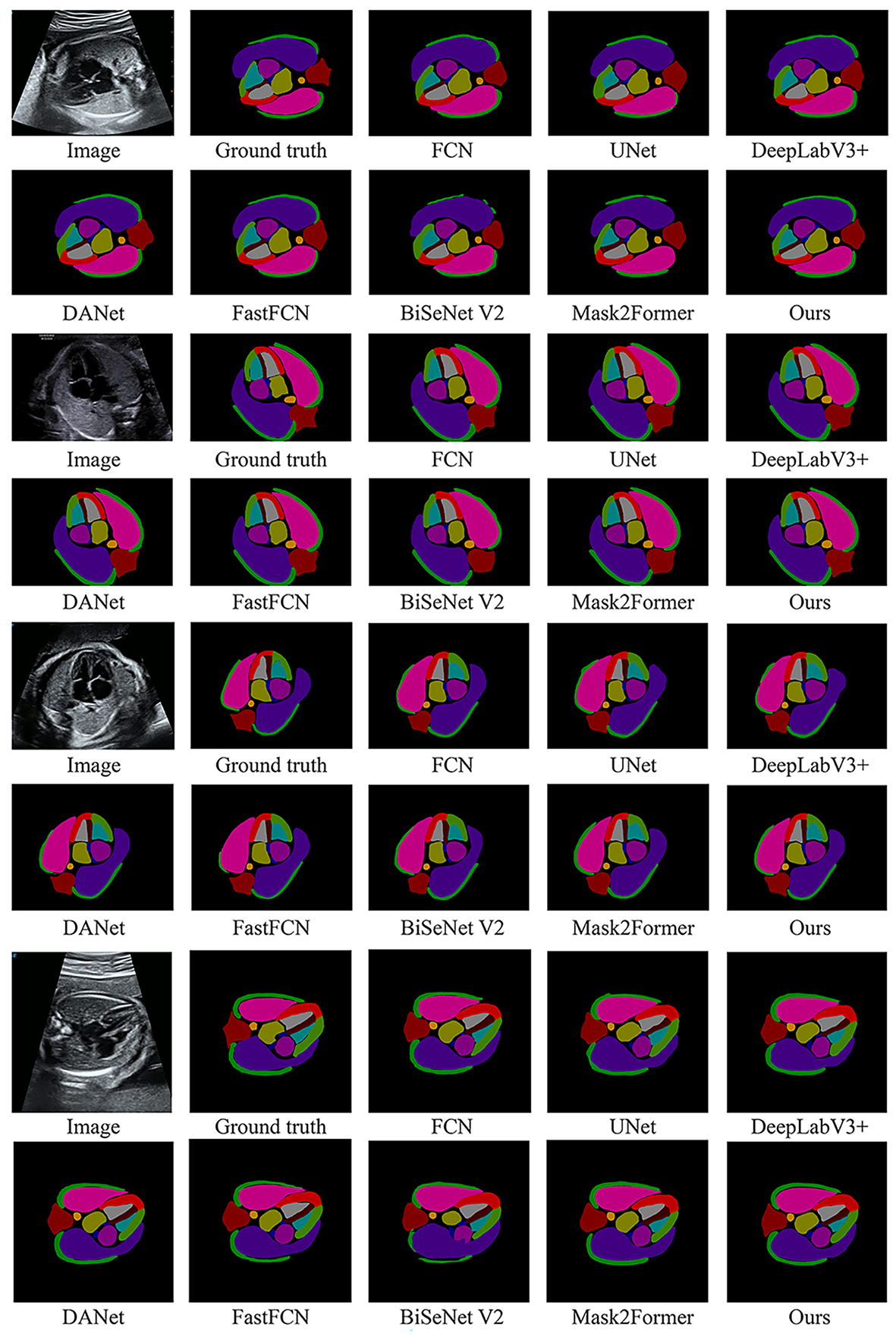}}
\caption{Visualization of segmentation results of different methods on the test set. The regions highlighted in Maroon, Green, Olive, Navy, Purple, Teal, Gray, Dark Red, Bright Red, Dark Olive Green, Dark Orange, Indigo and Deep Pink represent the SP, RIB, LA, IAS, RA, RV, LV, IVS, LVW, RVW, DAO, RL and LL, respectively.}
\label{fig2}
\end{figure}
As shown in Fig.~\ref{fig2}, we compared the proposed DCD model with several baseline models in terms of semantic segmentation results for anatomical structures on the fetal A4C view. Each row of images presents, from left to right, the original ultrasound image, the annotated ground truth segmentation, and the segmentation results of each model. From the visual results, it can be seen that the DCD model performs excellently in terms of segmentation accuracy and boundary coverage, with its results highly consistent with the ground truth. In contrast, some baseline models exhibit issues of under-segmentation and over-segmentation. BiSeNet V2, DANet, FastFCN, and DeepLabv3+ produce discontinuous results when segmenting the RiB anatomical structure, which may be due to the small and elongated shape of the RiB, making it difficult for the models to capture its complete boundaries and details. All methods perform poorly in segmenting the SP structure, but our approach achieves the best segmentation performance, clearly delineating the tip and details of the SP to the greatest extent. In summary, the DCD model demonstrates a significant advantage in the precise segmentation and detail handling of anatomical structures.

\section{Conclusion}
In this paper, we propose a semantic segmentation model named DCD designed for fetal ultrasound A4C view. The model is built upon the DeepLabv3+ architecture with several key enhancements. First, the standard ASPP is replaced with Dense ASPP to enhance the fusion of multi-scale information, allowing for more effective feature extraction across different scales and enhancing the model's performance in complex scenarios. Second, to further optimize feature representation, we introduce the CBAM attention mechanism into shallow features, enabling the model to focus more on critical regions and fine details, thereby improving segmentation accuracy. Experiments on an internal fetal cardiac ultrasound image dataset demonstrate that the DCD model demonstrates in multi-class anatomical structure segmentation for the fetal A4C view, particularly in the segmentation of fine structures, showing significant improvements compared to baseline models. In future work, we plan to extend the model to other types of fetal ultrasound images for the segmentation of additional key anatomical structures, aiming to enhance the model's applicability across different ultrasound images.
\subsubsection{\ackname}  
This research was partially supported by the National Natural Science
Foundation of China (62162054), the Guangxi Natural Science Foundation Program (2024JJA170233), the Basic Ability Improvement Project for
Young and Middle-aged Teachers in Guangxi, China (2024KY0694), the Key
Research Project of Wuzhou University (2023C004), the Wuzhou Science
and Technology Plan Project (202302036) and the National College Student Innovation and
Entrepreneurship Training Program Funded Projects (202411354042).
\bibliographystyle{spmpsci_unsrt}
\bibliography{svproc}

\begin{thebibliography}{10}
\providecommand{\url}[1]{{#1}}
\providecommand{\urlprefix}{URL }
\expandafter\ifx\csname urlstyle\endcsname\relax
  \providecommand{\doi}[1]{DOI~\discretionary{}{}{}#1}\else
  \providecommand{\doi}{DOI~\discretionary{}{}{}\begingroup \urlstyle{rm}\Url}\fi

\bibitem{In1}
Vullings, R.: Fetal electrocardiography and deep learning for prenatal detection of congenital heart disease.
\newblock In: 2019 Computing in Cardiology (CinC), pp. Page--1. IEEE (2019)

\bibitem{In2}
Taksande, A., Jameel, P.Z.: Critical congenital heart disease in neonates: a review article.
\newblock Current Pediatric Reviews \textbf{17}(2), 120--126 (2021)

\bibitem{In3}
Sood, E., Newburger, J.W., Anixt, J.S., Cassidy, A.R., Jackson, J.L., Jonas, R.A., Lisanti, A.J., Lopez, K.N., Peyvandi, S., Marino, B.S., et~al.: Neurodevelopmental outcomes for individuals with congenital heart disease: updates in neuroprotection, risk-stratification, evaluation, and management: a scientific statement from the american heart association.
\newblock Circulation \textbf{149}(13), e997--e1022 (2024)

\bibitem{In4}
Wang, Y., Ge, X., Ma, H., Qi, S., Zhang, G., Yao, Y.: Deep learning in medical ultrasound image analysis: a review.
\newblock IEEE Access \textbf{9}, 54,310--54,324 (2021)

\bibitem{In5}
Jeanty, P., Chaoui, R., Tihonenko, I., Grochal, F.: A review of findings in fetal cardiac section drawings: Part 1: The 4-chamber view.
\newblock Journal of Ultrasound in Medicine \textbf{26}(11), 1601--1610 (2007)

\bibitem{R11}
Lu, Y., Li, K., Pu, B., Tan, Y., Zhu, N.: A yolox-based deep instance segmentation neural network for cardiac anatomical structures in fetal ultrasound images.
\newblock IEEE/ACM Transactions on Computational Biology and Bioinformatics  (2022)

\bibitem{R16}
Zhou, Z., Lu, Y., Bai, J., Campello, V.M., Feng, F., Lekadir, K.: Segment anything model for fetal head-pubic symphysis segmentation in intrapartum ultrasound image analysis.
\newblock Expert Systems with Applications \textbf{263}, 125,699 (2025)

\bibitem{R13}
Chen, G., Tan, G., Duan, M., Pu, B., Luo, H., Li, S., Li, K.: Mlmseg: a multi-view learning model for ultrasound thyroid nodule segmentation.
\newblock Computers in Biology and Medicine \textbf{169}, 107,898 (2024)

\bibitem{R17}
Wang, Q., Zhao, D., Ma, H., Liu, B.: Fb-zwunet: A deep learning network for corpus callosum segmentation in fetal brain ultrasound images for prenatal diagnostics.
\newblock Biomedical Signal Processing and Control \textbf{104}, 107,499 (2025)

\bibitem{R15}
Wu, X., Tan, G., Pu, B., Duan, M., Cai, W.: Dh-gac: Deep hierarchical context fusion network with modified geodesic active contour for multiple neurofibromatosis segmentation.
\newblock Neural Computing and Applications pp. 1--16 (2022)

\bibitem{R18}
Hekal, A.A., Amer, H.M., Moustafa, H.E.D., Elnakib, A.: Automatic measurement of head circumference in fetal ultrasound images using a squeeze atrous pooling unet.
\newblock Biomedical Signal Processing and Control \textbf{103}, 107,434 (2025)

\bibitem{R12}
Pu, B., Lu, Y., Chen, J., Li, S., Zhu, N., Wei, W., Li, K.: Mobileunet-fpn: A semantic segmentation model for fetal ultrasound four-chamber segmentation in edge computing environments.
\newblock IEEE Journal of Biomedical and Health Informatics \textbf{26}(11), 5540--5550 (2022)

\bibitem{R14}
Zhao, L., Tan, G., Pu, B., Wu, Q., Ren, H., Li, K.: Transfsm: Fetal anatomy segmentation and biometric measurement in ultrasound images using a hybrid transformer.
\newblock IEEE Journal of Biomedical and Health Informatics  (2023)

\bibitem{R21}
Zhang, S., Peng, Z., Li, H.: Sau-net: Medical image segmentation method based on u-net and self-attention.
\newblock Acta electronica sinica \textbf{50}(10), 1--10 (2022)

\bibitem{R22}
Rajamani, S.T., Rajamani, K., Schuller, B.W.: A novel and simple approach to regularise attention frameworks and its efficacy in segmentation.
\newblock In: 2023 45th Annual International Conference of the IEEE Engineering in Medicine \& Biology Society (EMBC), pp. 1--4. IEEE (2023)

\bibitem{R23}
Zhou, W., Guan, G., Cui, W., Yi, Y.: Biasam: Bidirectional-attention guided segment anything model for very few-shot medical image segmentation.
\newblock IEEE Signal Processing Letters  (2024)

\bibitem{R24}
Tong, S., Zuo, Z., Liu, Z., Sun, D., Zhou, T.: Hybrid attention mechanism of feature fusion for medical image segmentation.
\newblock IET Image Processing \textbf{18}(1), 77--87 (2024)

\bibitem{R25}
Cao, Y., Cheng, Y.: Sacu-net: Shape-aware u-net for biomedical image segmentation with attention mechanism and context extraction.
\newblock IEEE Access  (2025)

\bibitem{DeepLabv3+}
Chen, L.C., Zhu, Y., Papandreou, G., Schroff, F., Adam, H.: Encoder-decoder with atrous separable convolution for semantic image segmentation.
\newblock In: Proceedings of the European conference on computer vision (ECCV), pp. 801--818 (2018)

\bibitem{UNet}
Ronneberger, O., Fischer, P., Brox, T.: U-net: Convolutional networks for biomedical image segmentation.
\newblock In: Medical image computing and computer-assisted intervention--MICCAI 2015: 18th international conference, Munich, Germany, October 5-9, 2015, proceedings, part III 18, pp. 234--241. Springer (2015)

\bibitem{Mask2Former}
Cheng, B., Misra, I., Schwing, A.G., Kirillov, A., Girdhar, R.: Masked-attention mask transformer for universal image segmentation.
\newblock In: Proceedings of the IEEE/CVF conference on computer vision and pattern recognition, pp. 1290--1299 (2022)

\bibitem{FCN}
Long, J., Shelhamer, E., Darrell, T.: Fully convolutional networks for semantic segmentation.
\newblock In: Proceedings of the IEEE conference on computer vision and pattern recognition, pp. 3431--3440 (2015)

\bibitem{DANet}
Fu, J., Liu, J., Tian, H., Li, Y., Bao, Y., Fang, Z., Lu, H.: Dual attention network for scene segmentation.
\newblock In: Proceedings of the IEEE/CVF conference on computer vision and pattern recognition, pp. 3146--3154 (2019)

\bibitem{FastFCN}
Wu, H., Zhang, J., Huang, K., Liang, K., Yu, Y.: Fastfcn: Rethinking dilated convolution in the backbone for semantic segmentation.
\newblock arXiv preprint arXiv:1903.11816  (2019)

\bibitem{BiSeNetV2}
Yu, C., Gao, C., Wang, J., Yu, G., Shen, C., Sang, N.: Bisenet v2: Bilateral network with guided aggregation for real-time semantic segmentation.
\newblock International journal of computer vision \textbf{129}, 3051--3068 (2021)

\end{thebibliography}
\end{document}